\documentclass[usenatbib,useAMS,times]{mn2e}
\usepackage{graphicx}

\usepackage{aas_macros}
\title[Optical variability of TVLM 513-46546]{Optical variability of the ultracool dwarf TVLM~513-46546: evidence for inhomogeneous dust clouds$^1$}

\author[]{S.\,P.\ Littlefair$^{1}$, V.\,S.\, Dhillon$^{1}$, 
T.\,R.\, Marsh$^{2}$, T.\, Shahbaz$^{3}$, \newauthor E.\,L.\, Mart\'{i}n$^{3,4}$, C. Copperwheat$^{2}$\\
$^1$Dept of Physics and Astronomy, University of Sheffield, Sheffield, S3 7RH, UK \\
$^2$Dept of Physics, University of Warwick, Coventry, UK\\
$^3$Instituto de Astrof\'{i}sica de Canarias, Tenerife, Spain\\
$^4$University of Central Florida, Dept. of Physics, PO Box 162385, Orlando, FL 32816-2385, USA\\}

\date{\center{\Large Submitted for publication in the Monthly Notices of the
Royal Astronomical Society \\ 
\vspace{.5cm} \today}} 

\begin{document}
\maketitle

\begin{abstract} 
We present multi-colour photometry of the M8.5V ultracool dwarf ``pulsar'' TVLM 513-46546 (hereafter TVLM 513) obtained with the triple-beam photometer {\sc ultracam}. Data were obtained simultaneously in the Sloan-$g'$ and Sloan-$i'$ bands. The previously reported sinusoidal variability, with a period of 2-hrs, is recovered here. However, the Sloan-$g'$ and Sloan-$i'$ lightcurves are anti-correlated, a fact which is incompatible with the currently proposed starspot explanation for the optical variability. The anti-correlated nature and relative amplitudes of the optical lightcurves are consistent with the effects of persistent dust clouds rotating on the surface of the star. In the absence of other plausible explanations for the optical variability of TVLM 513, it seems likely that dust cloud coverage combined with the rapid rotation of TVLM 513 is responsible for the optical variability in this object. However, crude modelling of a photosphere with partial dust cloud coverage shows that the anti-correlation can only be reproduced using cooler models than the literature temperature of TVLM 513. We suggest this discrepancy can be removed if more dust is present within the photosphere of TVLM 513 than theoretical model atmospheres predict, though a definitive statement on this matter will require the development of self-consistent models of partially dusty atmospheres.
\end{abstract} 


\section{Introduction}
\label{sec:introduction}
\alph{footnote} \protect\footnotetext[1]{Based on observations made at the European Southern Observatory, Paranal, Chile (ESO program 079.C-0686)} 
Brown dwarfs and very-low mass stars (collectively known as ultracool dwarfs, or UCDs) are strongly affected by the presence of dust in their photospheres. Dust absorbs elements from the gas phase, changing the opacity and metallicity of the atmosphere. Dust begins to form at temperatures corresponding to the transition between M and L spectral types, and becomes more prominent as the atmosphere cools; the presence of dust thus defines the L-dwarfs. Finally, when the dust clouds ``rain-out'' at lower temperatures still, they are responsible for the L-T transition. Thus understanding the formation, chemistry and atmospheric dynamics of dust is the central challenge facing theories of ultracool dwarf atmospheres \citep{burrows06}.

The presence of dust is also thought to affect the magnetic properties of ultracool dwarfs. A combination of an increasingly neutral atmosphere, and frequent collisions between charged particles and dust grains in the dense atmosphere results in the atmospheres of ultracool dwarfs having a high electrical resistivity \citep{mohanty02}. Thus, whilst strong quiescent and flaring radio emission reveals strong magnetic fields amongst the ultracool dwarfs \citep[e.g][]{berger01,burgasser05,hallinan07}, the predominantly neutral atmospheres may explain the relatively low levels of other activity indicators, such as H$\alpha$ emission and X-rays \citep[e.g][]{gizis00,west04}.

TVLM 513 (M8.5V) is an ideal object in which to study the interplay of magnetic fields with a cool, dense atmosphere. It is close by \citep[${\rm d}=10.6$\,pc; ][]{dahn02}, has known radio and H$\alpha$ activity, and is rapidly rotating \citep{mohanty03}. The radio emission of TVLM 513 is fascinating; observations from 1.4 to 8.5 GHz show the radio emission can switch states between highly polarised pulses with a 2-hr period \citep{hallinan07} to a fainter, quiescent state interrupted by stochastic flares \citep{berger07}. The periodic nature of the radio emission has led to this object being dubbed an ultracool dwarf ``pulsar''. Strong H$\alpha$ emission, modulated on the same 2-hr period, and a detection of X-ray emission \citep{berger07}, show that TVLM 513 is a magnetically active star, which is capable of supporting a chromosphere and a corona. This is despite its low effective temperature of $\sim$2300\,K \citep{dahn02}, roughly the temperature at which dust is believed to start forming within the photosphere \citep[see][and references within]{burrows06}. 

Optical I-band photometry of TVLM 513 revealed the same 2-hr periodicity as seen in the radio and H$\alpha$ emission; this was assumed to be due to starspots \citep{lane07}. However, optical variability in ultracool dwarfs can also be caused by dust clouds within the photosphere \citep[see][for a review]{bailer-jones02}. Simultaneous optical photometry in two or more bands can distinguish between variability caused by starspots \citep{rockenfeller06a} and dust clouds \citep{littlefair06}. We therefore obtained simultaneous, multi-colour photometry  using the fast photometer {\sc ultracam} \citep{dhillon07} to determine the cause of the optical variability in TVLM 513. The observations are described in
Section~\ref{sec:obs}, the results presented in Section~\ref{sec:results} and discussed in Section~\ref{sec:disc}, whilst in Section~\ref{sec:conc} we draw our conclusions.

\section{Observations \& Data Reduction}
\label{sec:obs}
For a duration of 2.7 hours on the night of June 9$^{\rm th}$ 2007, the M8.5V star TVLM 513 was observed simultaneously in the Sloan-$g'$ and $i'$ bands using {\sc ultracam} on the 8.2-m {\sc melipal} unit of the Very Large Telescope (VLT) at Paranal, Chile. Thanks to the frame-transfer CCDs employed in {\sc ultracam} we were able to obtain $\sim$10 second exposures with 24 msecs dead-time between frames. Sloan-$u'$ images were obtained simultaneously with the other colours, but the star is too faint to produce useful lightcurves in this band. The Sloan-$g'$ and $i'$ magnitudes of TVLM 513 are 20.9 and 16.1, respectively. Observing conditions were excellent; the night was photometric and the seeing stable around 0.6 arcseconds. Data reduction was carried out using the {\sc ultracam} pipeline data reduction software. The data were first bias subtracted and flat fielded. The flat field used was a median of $\sim$100 twilight frames taken on a blank piece of sky with the telescope dithering. Because the flat fields suffer from scattered light, large-scale trends were removed from the flat-field before use, by dividing the flat-field by a median-filtered version of itself. Since the position of the stars on the CCD remains constant throughout our observations, this will not affect our photometry. 

Extraction of target and reference star lightcurves was performed using a variable-sized aperture. The aperture size for both objects was set to 1.7 times the full-width at half maximum of the target stars spatial profile, as measured by a Gaussian fit.  Differential photometry was obtained by dividing by the flux of a nearby reference star. This implicitly assumes that the same proportion of the total flux falls outside the aperture for both target and reference stars. This is only true if the target and reference stars have the same point spread function (PSF). If the instrumental PSF of the target and reference stars differ, this introduces errors into the photometry. Worse still, since the actual PSF is the convolution of the instrumental PSF and the seeing profile, it is true that the {\em difference} in PSF between the target and reference (and hence error in the photometry) is a systematic function of seeing.  To minimise this source of error one can either use many reference stars, and fit the PSF as a function of chip position and time, or you can use a single reference star as close to the target as possible. Our images do not contain enough reference stars to implement the former approach, and so here we choose to implement the latter. The reference star chosen has g' and i' magnitudes of 17.2 and 16.6 and is 37 arcseconds from the target star.

Photometry for multiple reference stars was extracted and we have used this both as a ``sanity check" and as a measure of our photometric accuracy (although, for the reasons discussed above, this is likely to be an upper-limit). The photometric accuracy we determine is 3 mmag for the $i'$-band and 7 mmag for the $g'$-band. This is smaller than the sinusoidal variability reported in section~\ref{sec:results}, giving us confidence that this variability is real. Indeed, the sinusoidal variability is recovered, albeit with more noise, if other reference stars are used to produce the differential photometry.

\section{Results} 
\label{sec:results}

\begin{figure*}
\begin{center}
\includegraphics[scale=0.5,trim=0 0 0 0,angle=-90]{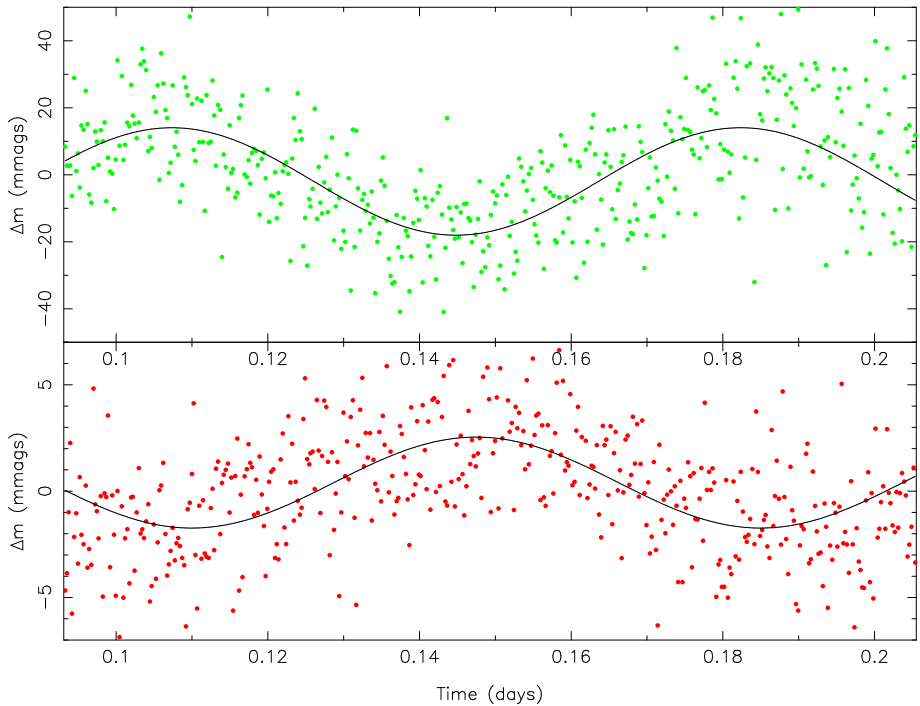}
\caption{Sloan-$g'$ and $i'$ lightcurves of TVLM 513. The change in brightness of the target star (mmags), is plotted against time in days. The data have been binned into bins of 20 second duration. The Sloan-$g'$ lightcurve is plotted in green in the top panel, whilst the bottom panel shows the Sloan-$i'$ lightcurve, plotted in red.  Note the vertical scales for the $g'$-band lightcurve is much larger than that for the $i'$-band lightcurve. Also shown are sinusoidal fits to the data}
\label{fig:lightcurves}
\end{center}
\end{figure*}
The {\sc ultracam} photometry is shown in Figure~\ref{fig:lightcurves}.  Both the $g'$ and $i'$ lightcurves exhibit sinusoidal variability with a period of around two hours. To assess the significance of this variability, 1000 copies of the data were made with the time-ordering of the data points randomised. Lomb-Scargle periodograms of these copies were computed and compared to the periodograms of the original data. None of the randomised datasets showed power in the periodograms as large as the peak power of the actual data, and so the false alarm probability for the periodic variability is less than 0.1 per cent in both bands. Errors in the periods were computed using a bootstrapping technique. We computed 1000 copies of the data using bootstrapping with replacement. Periods for these copies were determined from the peak of a Lomb-Scargle periodogram, and the scatter in these values used as an estimate of our period error. We find periods of $2.0 \pm 0.5$ hours for the $g'$-band and $2.4\pm0.4$ hours for the $i'$-band. This confirms the periodic variability seen in the I-band lightcurves of TVLM 513 \citep{lane07}, which is also seen in the equivalent width of H$\alpha$ emission \citep{berger07}. 

The RMS variability in the $g'$-band is about 1.5 per cent. RMS variability in the $i'$-band is 0.15 per cent. The $i'$-band variability is thus much smaller than that reported in the I-band by \cite{lane07}, which was nearer 0.8 percent. The most striking fact about the optical variability is that the $g'$ lightcurve and $i'$ lightcurves are in anti-phase. A discrete cross-correlation function \citep{edelson88} yields a phase shift between the two lightcurves of $0.50\pm0.05$. It is not easy to reconcile this observation with the claim that the broadband optical variability of TVLM 513 is due to starspots \citep{lane07}; since starspots cause a dimming of the star at all wavelengths, starspot-induced variability would produce $g'$ and $i'$ lightcurves which are in phase. We therefore turn our attention to other possible causes of the optical variability.

\subsection{Starspots and Faculae}
Whilst starspots alone cannot reproduce the optical variability, it is possible in principle to explain it using a combination of starspots and faculae. Because the faculae are hotter than the surrounding photosphere, we can conceive of a situation in which the effect of cool starspots dominates in the $i'$-band, whilst the faculae have a dominant effect in the $g'$-band. In these circumstances, an excess of spots would cause a dimming in the $i'$-band. Due to the corresponding excess of faculae, this would be accompanied by a brightening in the $g'$-band.

In the absence of constraints on spot and faculae properties in ultracool dwarfs, we cannot rule out this scenario entirely. However, we note the following points, which make this explanation unlikely. First, to ensure that the effect of spots dominates in the $i'$-band, whilst the effects of faculae dominate in the $g'$-band requires considerable fine-tuning of the relative areas of spot and faculae coverage. We find, that for plausible spot and faculae temperatures of $T_s = 1800$\,K and $T_f = 2400$\,K, respectively, the ratio of faculae area to spot area must lie between 1.4 and 2.2. This fine tuning of relative areas makes this scenario unlikely a-priori. Furthermore, the $g'$ and $i'$ lightcurves are observed to vary in anti-phase to a good level of accuracy (see Figure~\ref{fig:lightcurves}). This requires additional fine-tuning of the {\em spatial} arrangement of the spots and faculae, as the contrast of faculae is greatest at the limb, whereas spot contrast is roughly independent of limb position \citep{gondoin08}. For these reasons, we consider an explanation of the optical variability in terms of spots and faculae unlikely.

\subsection{Chromospheric emission}
\begin{figure}
\begin{center}
\includegraphics[scale=0.3,trim=0 0 0 0]{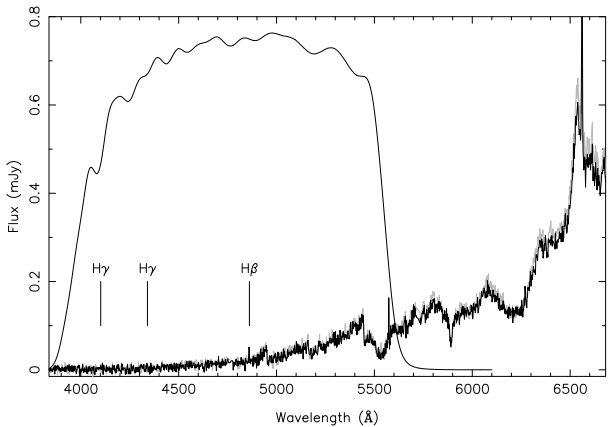}
\caption{Gemini spectra of TVLM 513 from \protect\cite{berger07}. The spectra in high and low-Balmer emission states are shown. The high Balmer emission state is plotted in black, with the low-Balmer emission state shown in light grey. Both spectra have been normalised by the peak flux of the high Balmer emission state spectrum. The {\sc ultracam} response curve in the $g'$-band is overplotted. A weak H$\beta$ emission line is visible in the high-state spectrum.}
\label{fig:spectrum}
\end{center}
\end{figure}
Anti-correlated variability in the Sloan-$g'$ and $i'$ bands is possible with a combination of starspots and chromospheric emission. Whilst the Ca\,HK lines lie outside the $g'$ and $i'$ bands,  several of the higher-order Balmer lines fall in the $g'$ band. Thus a more active region of the star might show lower $i'$-band flux (due to the dimming caused by starspots), but a higher $g'$-band flux (due to the contribution to the $g'$-band from Balmer emission lines). However, the spectrum of TVLM 513 appears to rule out this conjecture (Figure~\ref{fig:spectrum}). From the Gemini spectra of \cite{berger07} we can see that the higher-order Balmer lines contribute very little to the $g'$-band flux, even in the high state. In fact, Balmer emission only contributes an additional 0.3 per cent to the $g'$-flux, so it is unable to account for the 3 per cent peak-to-peak variability seen in the $g'$-lightcurve. Since the H$\alpha$ equivalent widths in the \cite{berger07} spectra span the range of observed H$\alpha$ equivalent widths in the literature for TVLM 513, it is likely that the Balmer emission in these spectra adequately reflects the Balmer emission at the time of our observations, and so we judge it unlikely that chromospheric emission is responsible for the optical variability.

\subsection{Magnetospheric Emission}
The radio emission of TVLM 513 is highly variable in nature. The radio emission seems to change between the bright pulsations with 2-hr period reported by \cite{hallinan07} to a fainter, quiescent, state interrupted by stochastic flares \citep{berger07}. There is some debate as to whether the radio emission results from an electron cyclotron maser (ECM) emission mechanism \citep{hallinan06}, or the same gyrosynchrotron emission which powers the radio emission in warmer M-dwarfs \citep{berger07}. As a narrow-band emission mechanism, ECM emission will not contribute to the optical flux, and so here make the conservative assumption that the radio emission is caused by gyrosynchroton radiation.

It is beyond the scope of this letter to determine whether gyrosynchrotron radiation can provide sufficient flux at optical wavelengths to cause the variability seen here. However, since gyrosynchrotron radiation produces a broad continuum, it is not likely that it is {\em directly} responsible for the anti-correlation between the $g'$ and $i'$ band variability. It is possible to construct a viable physical model whereby the $i'$-band variability is caused by starspots, and the $g'$-band variability is caused by variations in the gyrosynchrotron radiation. Such a model can plausibly produce anti-correlated variability between the two bands, if areas of the star with an excess of starspots are associated with brighter gyrosynchrotron radiation. The sinusoidal nature of the $g'$-band lightcurve argues against such a model, however, as the emission in the radio is not observed to vary sinusoidally. We therefore believe it is unlikely that magnetospheric emission, or a combination of magnetospheric emission and starspots, is responsible for the optical variability.

\subsection{Dust clouds}
\begin{figure}
\begin{center}
\includegraphics[scale=0.4,trim=50 0 0 80,angle=90]{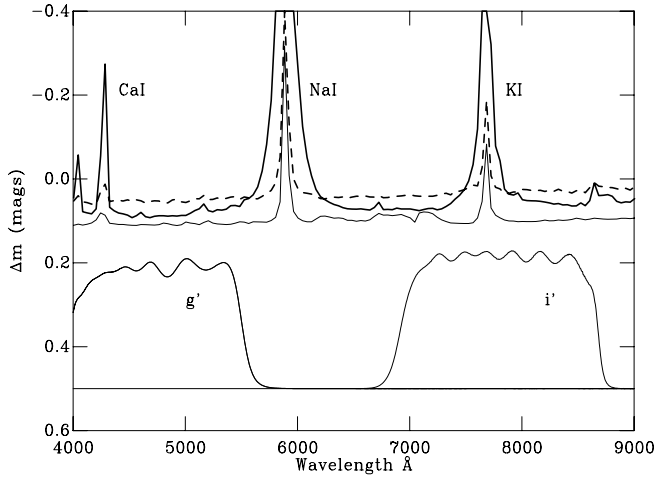}
\caption{Top: Calculations of the {\em change} in spectrum of an ultracool dwarf, due to the appearance of a dust cloud
that covers 10 per cent of the stellar surface. The calculations use the synthetic spectra of \protect\cite{allard01} - see text for details. The thick solid line shows a model with T$_{\rm eff}=1900$\,K, whilst the thick dashed line shows a model with T$_{\rm eff}=2300$\,K. The thin solid line shows a hybrid model in which dust forms within a 1900\,K starspot on a 2300\,K photosphere. Bottom: Also shown are the bandpasses of the Sloan-$g'$ filter and the Sloan-$i'$ filter.}
\label{fig:model}
\end{center}
\end{figure}
The presence of dusts clouds can, in principle, explain the optical variability seen in TVLM 513. Consider the formation of a dust cloud in a largely dust-clear photosphere. The cloud increases the continuum opacity, whilst at the same time the condensation in dust clouds of gas-phase alkali-metals reduces the opacity in spectral regions dominated by alkali absorption. Since the Sloan-$g'$ band is dominated by continuum opacity, whilst the Sloan-$i'$ band is dominated by molecular absorption, the formation of a dust cloud will reduce $g'$-band flux, and simultaneously increase the $i'$-band flux. Following \cite{bailer-jones02} we create a toy model of this process by combining the COND and DUSTY\footnote{Both COND and DUSTY models include dust for the purposes of calculating the chemistry, but COND models assume the dust settles below the photosphere, whilst DUSTY models assume the dust remains in the visible part of the atmosphere} models of \cite{allard01} in the ratio 1:9, and comparing the resulting spectrum with a 100 per cent COND model of the same effective temperature. We would like to stress that this model is only meant as an illustration, and is far too crude to attempt a detailed analysis such as lightcurve modelling, or derivation of cloud coverage. Such an analysis would require more sophisticated modelling which properly and self-consistently treated the dust content of cloudy and "cloud-free" regions of the photosphere. Examples of the spectral changes resulting from our toy model are shown in Figure~\ref{fig:model} for effective temperatures of 1900\,K and 2300\,K, as compared to the effective temperature for TVLM 513 of $\sim$2300 K \citep{dahn02}.

The models in Figure~\ref{fig:model} show that both the $g'$ and $i'$ bands show spectral changes due to resonance lines and changes in the continuum brightness. Whether the star gets brighter or fainter in a given band depends on whether line or continuum changes dominate. We determine this by folding the spectral changes in Figure~\ref{fig:model} through the $g'$ and $i'$ bandpasses.  For the 2300\,K model we find the continuum change dominates in both bands; the warmer model thus predicts {\em correlated} variability in the $g'$ and $i'$ bands. By contrast, in the 1900\,K model, the resonance lines dominate the $i'$-band whilst the continuum dominates the $g'$-band. The cooler model thus predicts the observed {\em anti-correlated} $g'$ and $i'$ variability. Therefore, only by using significantly cooler synthetic spectra than appropriate for TVLM 513 can we obtain the observed anti-correlation between $g'$ and $i'$. Figure~\ref{fig:model} shows why this should be the case. Whilst the models at 2300\,K and 1900\,K produce spectral changes of similar shape, it is clear that, in the 2300\,K model, the change in the alkali metal lines is much smaller. This is presumably a natural consequence of the fact that the amount of dust formed in the warmer model is itself much smaller than in the cooler one. Therefore, if the models underestimate the amount of dust that forms in a 2300\,K photosphere, it would be possible to reproduce the observed anti-correlation variability in $g'$ and $i'$ using models of the appropriate effective temperature. One might suspect this problem can be overcome using a ``hybrid model" in which dust clouds form in the cooler regions of the photosphere created by starspots. Figure~\ref{fig:model} shows that this is not the case. In the figure we show such a model, with a dusty, 1900\,K starspot covering 10 per cent of the surface of a dust-free 2300\,K photosphere. The starspot introduces such strong variability that it overwhelms the effects of the dust clouds, and this model produces correlated variability in the $g'$ and $i'$ bands.

\section{Discussion}
\label{sec:disc}

Based upon a lack of plausible alternative explanations for the optical variability in TVLM 513, we suggest that the most likely cause of the variability we observe is the rotation of a dust cloud covering a significant fraction of the stellar photosphere. This explanation is not without its problems. Crude modelling of a photosphere with partial dust cloud coverage shows that the observed optical variability can only be reproduced using models which are too cool by some 400\,K.  This may be an indication that  the models of \cite{allard01} underestimate the amount of dust that has formed in the photosphere of TVLM 513. Alternatively, this discrepancy could be a consequence of our simplistic modelling process; a definitive statement on this matter awaits the arrival of self-consistent models of the emission from partially dusty atmospheres.

Based upon the strong magnetic field inferred from radio observations of TVLM 513, and the strong H$\alpha$ emission seen in this object, it is perhaps surprising that the optical variability is not dominated by starspots, as was previously assumed. Indeed, optical multi-band photometric monitoring by \cite{rockenfeller06a} shows that starspots can cause periodic variability in objects as late as M9V. In this context, it is worth mentioning that the I-band variability reported by \cite{lane07} is four times larger than the variability reported here in $i'$. Such large amplitudes are difficult to explain using a dust cloud model, and may indeed be due to starspots. Thus, there is some evidence that the nature of the optical variability in TVLM 513 might not be stable, being at times dominated by dust clouds, at other times by starspots.  Further 
multi-wavelength observations will be needed to see if the nature of the optical variability is linked to changes in the radio behaviour \citep{hallinan07,berger07} of this remarkable object.

\section{Conclusions}
\label{sec:conc}

We present simultaneous monitoring of the M8.5V star TVLM 513 in the Sloan-$g'$ and $i'$ bands. Both bands
show sinusoidal variability on the $\sim$2-hr rotation period. The $g'$ and $i'$ band lightcurves are in anti-phase, a fact which is incompatible with the optical variability being due to starspots. Without a plausible alternative explanation
for this behaviour, we conclude that the optical variability in TVLM 513 is caused by the presence of a dust cloud in a predominantly dust-free photosphere. The optical variability is consistent with a dust cloud covering a significant fraction of the photosphere, but only if cooler stellar models than appropriate are used. This could be a consequence of the crudity of our modelling, or it may imply there is more dust in the atmosphere of TVLM 513 than the stellar atmosphere models of \cite{allard01} predict. The $i'$-band variability reported here is a factor of four smaller than previously reported by \cite{lane07}. Since such a high amplitude of variability is difficult to reconcile with dust cloud models, we suggest that the optical variability of TVLM 513-46546 can change in nature. It is unclear if this is related to the observed changes in radio emission from this object.

\section*{\sc Acknowledgements}
We thank Edo Berger for kindly providing his Gemini spectra in electronic form. The authors also thank the anonymous referee, whose comments greatly improved the quality of this manuscript. ULTRACAM is supported by PPARC grant PP/D002370/1. SPL is supported by an RCUK fellowship and PPARC grant PPA/G/S/2003/00058. TRM and CC are supported under grant ST/F002599/1 from the Science and Technology Facilities Council.This research has made use of NASA's Astrophysics Data System Bibliographic Services.

\bibliographystyle{mn2e}
\bibliography{abbrev,refs,refs2}

\end{document}